\documentclass[proceedings]{JHEP} 

\usepackage{epsfig}			

\newbox\mybox
\newcommand\fverb{\setbox\mybox=\hbox\bgroup\verb}
\newcommand\fverbdo{\egroup\medskip\noindent\fbox{\unhbox\mybox}\ }
\newcommand\fverbit{\egroup\item[\fbox{\unhbox\mybox}]}

\font\beeg=cmr17 scaled 1600		
\newcommand\init[1]{\setbox\mybox=\hbox{{\beeg #1}~}%
		   \noindent\global\hangindent=\wd\mybox\global\hangafter-2%
		   \sc\smash{\llap {\lower 13.2pt \box\mybox}}}

\title{QCD Sum Rules for Heavy Flavors}

\author{V.M. Braun \\
        Institut f\"ur Theoretische Physik, Universit\"at
        Regensburg, D-93040 Regensburg, Germany\\
	E-mail: \email{Vladimir.Braun@physik.uni-regensburg.de}}

\conference{Heavy Flavours 8, Southampton, UK, 1999}

\abstract{
I give a short summary of QCD sum rule results for hadrons involving a heavy 
quark, with emphasis on recent developments. 
}

\begin{document} 

\maketitle 


\section{Decay constants: $f_B$ {\em etc.}}

QCD sum rules for D-meson and B-meson decay constants were 
among the first applications of this approach \cite{SVZ}. Early estimates 
\cite{Shuryak82} have given $f_B\sim 100$~MeV and $m_b\sim 4.8$~GeV.
In 1983 Aliev and Eletsky \cite{AE83} carried out the first quantitative 
analysis incorporating  radiative corrections and relativistic effects 
(power-like $1/m_b$ corrections) with the result $f_B=130\pm 30$~MeV. 
This old analysis remains generally valid today and subsequent refinements 
concentrated on two issues.
  
First, it was noticed by several authors \cite{scatter} 
that the QCD sum rule result for $f_B$ is strongly sensitive to the 
value of the $b$-quark mass. More precisely, the sensitivity is
mainly to the difference $m_B-m_b$ rather than $m_b$ itself, but anyhow
lowering $m_b$ by 200 MeV produces an increase in $f_B$ by $\sim 50$~MeV. 
Another way to formulate the same problem is that the QCD sum rule
does not have enough accuracy to predict 
the values of $m_b$ and  $f_B$ {\em simultaneously}. Hence, $m_b$ 
has to be taken from elsewhere as an input parameter. 

Second, using the heavy quark effective theory one was able to 
resum \cite{BG92,BBBD92,N92}
all leading logarithmic contributions to the sum rule of the type 
$(\alpha_s(m_b)\ln m_b/\mu)^k$. The net effect of this resummation
is that the strong coupling in the radiative correction has to be 
taken at a low scale of order 1 GeV rather than $m_b$, as was assumed in 
\cite{AE83}. Combined with higher $\alpha_s$ values accepted nowadays,
this resummation induces an increase in $f_B$ by $30-60$~MeV, depending 
on further details.

The issue of the $b$-quark mass has recently been reexamined 
using new results \cite{BenekeHF8} on the resummation of Coulombic 
corrections to heavy-heavy correlation functions.
Several independent studies (see Table.~1)
all result in rather large quark mass values, not inconsistent with 
$m_b=4.8$ used in early QCD sum rule estimates. 
\TABLE[ht]{
\renewcommand{\arraystretch}{1.2}
\addtolength{\arraycolsep}{1pt}
$
\begin{array}{|c|c|c|c|}
\hline & m_b & \overline{m}_b(\overline{m}_b) & \mbox{Remarks}\\ 
\hline
\mbox{BS99 \cite{BS99}} & 4.97\pm 0.17 & 4.25\pm 0.08 & \mbox{Sum rules}  \\
\mbox{PP98 \cite{PP98}} & 4.80\pm 0.06 & \mbox{--} & \mbox{Sum rules} \\
\mbox{H98 \cite{H98}} & 4.88\pm 0.10 & 4.20\pm 0.06 & \mbox{Sum rules}  \\
\mbox{MY98 \cite{MY98}} & \mbox{--}  & 4.20\pm 0.10 & \mbox{Sum rules}  \\
\hline
\mbox{JP97/98 \cite{JP97/98}} & 4.60\pm 0.02 & 4.19\pm 0.06  &  
\mbox{Sum rules, no resummation}\\
\hline
\mbox{PY98 \cite{PY98}} & 5.00^{+0.10}_{-0.07} & 4.44^{+0.03}_{-0.04} & 
\Upsilon(\mbox{1S})\mbox{ mass} \\
\mbox{GGRM99 \cite{GGRM99}} & \mbox{--}  & 4.26\pm 0.08  & 
\mbox{lattice HQET} \\
\hline
\end{array}
$
\renewcommand{\arraystretch}{1}
\addtolength{\arraycolsep}{-1pt}
\caption{
Summary of recent determinations of the $b$-quark mass. 
The table is adapted from \cite{BSS99rep} 
and updated using \cite{BS99,GGRM99}.
}
}
The possibility 
to pin down the $b$-quark mass with good precision is encouraging
as it means  that the sensitivity of sum rules on $m_b$ is a less 
serious problem  as thought before. To match the accuracy 
of NNLO NRQCD calculations, one may try to
calculate the $O(\alpha_s^2)$ corrections to the sum rule  
which is possible by existing  methods.  
The BLM-type estimate \cite{BBB95} indicates that such correction is 
probably not large.  
One has to have in mind, however,  that accuracy of 
QCD sum rules cannot be consistently improved by calculated higher 
order corrections and such calculations are only meaningful as to
exclude (unexpected) large effects.

To summarize, I give my personal ``weighted average''
\begin{equation}
    f_B = 160 \pm 30~{\rm MeV}\,, 
\label{fB}
\end{equation}
which is in agreement with other estimates \cite{Dominguez93,Babar},
is stable for may years and, unfortunately, not improvable.
Note that this value is consistent with the old result of Aliev-Eletsky 
from 1983 and the error is actually not reduced despite significant 
effort that had been invested ever since.  

The coupling $f_D$ historically attracted less attention.
Aliev and Eletsky \cite{AE83} have obtained $f_D=160$~MeV with claimed 
accuracy of order 20\%. The dependence on $c$-quark mass happens to be 
somewhat  milder in this case. The recent updates typically fall in the range
 \cite{Dominguez93, KRWY99} 
\begin{equation}
  f_D = 190 \pm 30~{\rm MeV}\,,
\end{equation}
where the small increase is mainly due to a higher value of $\alpha_s$.

Calculation of $SU(3)$-breaking effects in the QCD sum rules is not 
easy because several effects tend to compensate each other.
This difficulty is reflected in the rather large errors  \cite{Babar}:
\begin{eqnarray}
 f_{D_s}/f_D &=& 1.19 \pm 0.08\,,
\nonumber\\
 f_{B_s}/f_B &=& 1.19 \pm 0.08\,.
\label{SU3}
\end{eqnarray}
Another result is (see e.g. \cite{BBKR95,KRWY99})
\begin{eqnarray}
 f_{D^*}/f_D &=& 1.40 \pm 0.15\,,
\nonumber\\
 f_{B^*}/f_B &=& 1.10\pm 0.08\,. 
\end{eqnarray}
Finally, with the advent of the Heavy Quark Effective Theory (HQET)
it has become customary to present relativistic corrections \cite{AE83}
to $f_B$ in the form of a  series 
expansion in powers of the heavy quark mass:
\begin{equation}
  f_B = \widehat{C}(m_b)F_{\rm stat}\left[1-\frac{A}{M_b}+\ldots\right].  
\end{equation} 
Here $F_{\rm stat}$ is a universal nonperturbative constant, 
$\widehat{C}(m_b)$ can be calculated perturbatively and the power
corrections can be related to matrix elements of higher dimension
HQET operators. The QCD sum rules give very stable predictions 
 \cite{ES92,BBBD92,Ball94}
\begin{equation}
 A = 0.9\pm 0.2~{\rm GeV}\,.
\end{equation} 

\section{HQET parameters $\bar{\Lambda},\lambda_1,\lambda_2$}

QCD sum rules have a well-defined heavy quark limit and 
can be used to estimate HQET parameters, see \cite{Nreview} for 
the relevant definitions.
For example, one obtains  for the difference between meson and quark masses
\cite{BBBD92,N92}
\begin{equation}
  \bar\Lambda \simeq 400 - 500~{\rm MeV}\,.
\end{equation}
This value is consistent with $m_b$ determinations although the accuracy is 
not competitive. 
More interestingly, the sum rule for the chromomagnetic interaction parameter 
$\lambda_2$ comes out to be rather stable. Translated into the prediction
 for the mass splitting  between vector and pseudo\-sca\-lar mesons, 
the result reads \cite{N92,BB94,N96} 
\begin{equation}
 (m_V^2-m_P^2)^{\mu=m_b} = (0.46\pm 0.14)~{\rm GeV}^2\,, 
\end{equation} 
which compares very well with the experimental value 
$m_{B^*}^2-m_B^2 \simeq 0.48$~ GeV$^2$. A technical remark is that 
all sum rule results in HQET apear to be strongly correlated with 
the value of $\bar\Lambda$ which is just another manifestation of 
the sensitivity of ``ordinary'' sum rule on the value of $m_b$. 
To reduce this dependence, and also to moderate radiative 
corrections, one often attempts to consider ratios
of sum rules in which such effects tend to cancel.

The most interesting quantity is $-\lambda_1$, the $b$-quark kinetic
energy in the B-meson. Knowledge of $\lambda_1$ is crucial for the 
accuracy of $V_{cb}$ determinations from inclusive B-decays using OPE.
The situation with $\lambda_1$ determination from QCD sum rules
was controversial for some time, with two calculations reporting 
conflicting results: $-\lambda_1 = 0.5\pm 0.2$~GeV$^2$~\cite{BB94} and
$-\lambda_1 = 0.1\pm 0.05$~GeV$^2$ \cite{N96}. By now, the origin 
of this discrepancy is well understood \cite{BSU97}, but the remedy 
is not found.
The difficulty can be traced to contributions of nondiagonal 
matrix elements
$\sim\langle B|\bar h (D_\perp^2)h|B'\rangle$ where 
$B'$ is a certain excited pseudoscalar meson state, which are 
difficult to disentangle from the ``diagonal'' contribution of interest 
within the QCD sum rule framework. Several toy-model calculations 
using quantum-mechanical examples \cite{BL98} strongly suggest that 
such contributions are present in both sum rules \cite{BB94} 
and \cite{N96} and have opposite signs. In quantum mechanics 
it is always possible to add  the two sum rules 
analogous to \cite{BB94,N96} 
with some weight (fixed by the virial 
theorem) such that the nondiagonal transitions cancel exactly. In QCD
the corresponding weight  is not known. Therefore, for the time being 
I prefer to take the results of \cite{BB94,N96} 
as providing for the 
upper and the lower boundary, respectively, 
leading to a conservative average
\begin{equation}
 -\lambda_1 = 0.35\pm 0.20~{\rm GeV}^2\,.
\end{equation}
Doing better is an open problem. An unrelated difficulty is that
the value for $-\lambda_1$ may depend rather strongly on its 
precise definition, in
particular whether or not a Lorentz-invariant UV cutoff is used. 

\section{$B\to D,\ldots$ form factors and Isgur-Wise functions}

In the period 1989 -- 1992 several sum rules have been derived 
and studied for the semileptonic decays of beauty to charm, 
using the standard QCD sum rule maschinery and finite values
for heavy quark masses \cite{OS89,BG90,CNOP91}. In later works
the accent has shifted to incorporate the heavy quark expansion
and calculate the Isgur-Wise functions instead of the form factors 
themselves. Such approach has an advantage that the heavy quark symmetry
is incorporated analytically rather than numerically, but at the same 
time raises concerns on  validity of the $1/m_c$ expansion. 

A lot of effort went into  calculation of  the celebrated Isgur-Wise function
$\xi(y)$ and its slope $\rho = \xi'(y=1)$ as providing main input  
for the $V_{cb}$ determinations. The leading-order sum rules
 \cite{Rad91,N92,Ball92,BS93} have later been complemented by calculation of 
the radiative corrections \cite{IWradcor}. The result is
\begin{equation}
 \rho = 0.7 \pm 0.25
\end{equation} 
and probably cannot be improved. The discussion around this sum rule was
actually beneficial for the general development of the sum rule approach, 
allowing to clarify certain important points about continuum subtraction,
see \cite{BS93} for details. 

QCD sum rules have also been derived for the Isgur-Wise functions
$\tau_{1/2}(y)$ and $\tau_{3/2}$ which govern semileptonic transitions 
to positive parity states $\frac{1}{2}^- \to \frac12^+, \frac32^+$
and contribute e.g. to the Bjorken inequality 
\begin{equation}
  \rho > \frac14 + |\tau_{1/2}(1)|^2 + |\tau_{3/2}(1)|^2\,.
\end{equation}  
The sum rule for $\tau_{3/2}$ is only available to leading order \cite{CNP92}
while for $\tau_{1/2}$ the radiative corrections are included \cite{CFP98}.
The  results are: 
\begin{eqnarray}
 \tau_{1/2}(1) = 0.35\pm 0.08 &,& \rho_{1/2}^2= 2.5\pm 1.0~
  \mbox{\protect{\cite{CFP98}}}\,,
\nonumber\\
 \tau_{3/2}(1) \sim 0.3  &,& \rho_{3/2}^2 \sim 0.9~
  \mbox{\protect{\cite{CNP92}}}\,.
\end{eqnarray}
For illustration of the  accuracy, I show one typical plot: 
\FIGURE{\epsfig{file=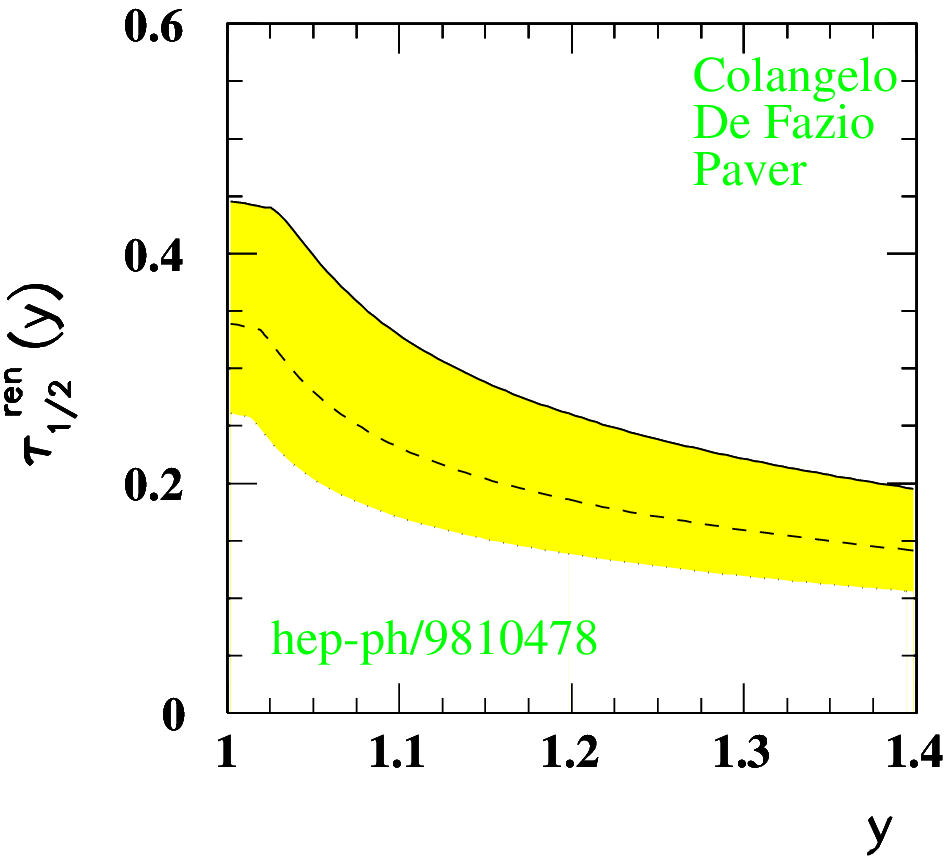,width=6cm}%
        \caption{The Isgur-Wise function $\tau_{1/2}$. Figure taken from
\protect{\cite{CFP98}.} 
}
	\label{tau12}}

Finally, certain results are available for subleading Isgur-Wise 
functions $\chi_{1,2,3}(y)$ and $\xi_3(y)$ \cite{sub}. 
See also \cite{HLD99}.

\section{Semileptonic D decays}

Calculations of the form factor values at $q^2=0$ have a long history
\cite{AEK84}-\cite{Ball93} and all are rather old. 
The results are \cite{BBD91,Ball93}
\begin{eqnarray}
   f_+^{D\to \pi}(0) &=& 0.50\pm 0.15\,,\nonumber\\
    f_+^{D\to K}(0) &=& 0.60\pm 0.15
\end{eqnarray}
and
\begin{eqnarray}
A_1^{\rho}(0) &=& 0.50\pm 0.2\,, \nonumber\\ 
A_1^{K^*}(0)  &=& 0.50\pm 0.15\,, \nonumber\\
A_2^{\rho}(0) &=& 0.40\pm 0.1\,, \nonumber\\
A_2^{K^*}(0)  &=& 0.60\pm 0.15\,, \nonumber\\ 
V^{\rho}(0)   &=& 1.0\pm 0.2 \,,\nonumber\\
V^{K^*}(0)    &=& 1.1\pm 0.25\,.
\end{eqnarray}
The quoted values for $f_+^{D\to \pi}(0)$ and $f_+^{D\to K}(0)$ are 
somewhat too low, to my opinion, and result from using a too small
value for the quark condensate. The other entries are less affected
by this choice.
   
QCD sum rule predictions for the $q^2$ dependence of the form factors 
\cite{BBD91} have been the most interesting.
While the shape of the form factors $f_+$ and $V$ turned out to be consistent
with vector dominance, the axial-vector form factors $A_1$ and $A_2$ came 
out almost flat. Such behaviour was unexpected and initiated a lively 
discussion which continues nowadays. It is worth while to mention that 
sum rule calculations of the form factors in the physical 
region  $q^2>0$ are technically nontrivial because of subtleties 
in the construction of  double dispersion 
relations in presence of Landau thresholds, see \cite{BBD91} for the details. 

In addition, the form factor 
$f_+^{D\to \pi}(q^2)$ was analysed using light-cone sum rules 
(LCSR) in \cite{BBKR95}
(see the next section for a detailed description of this approach).
Very recently, the NLO corrections  to this sum rule 
have been calculated \cite{K}, with preliminary results shown 
in Fig.~\ref{fplusDpi}:

\FIGURE{\epsfig{file=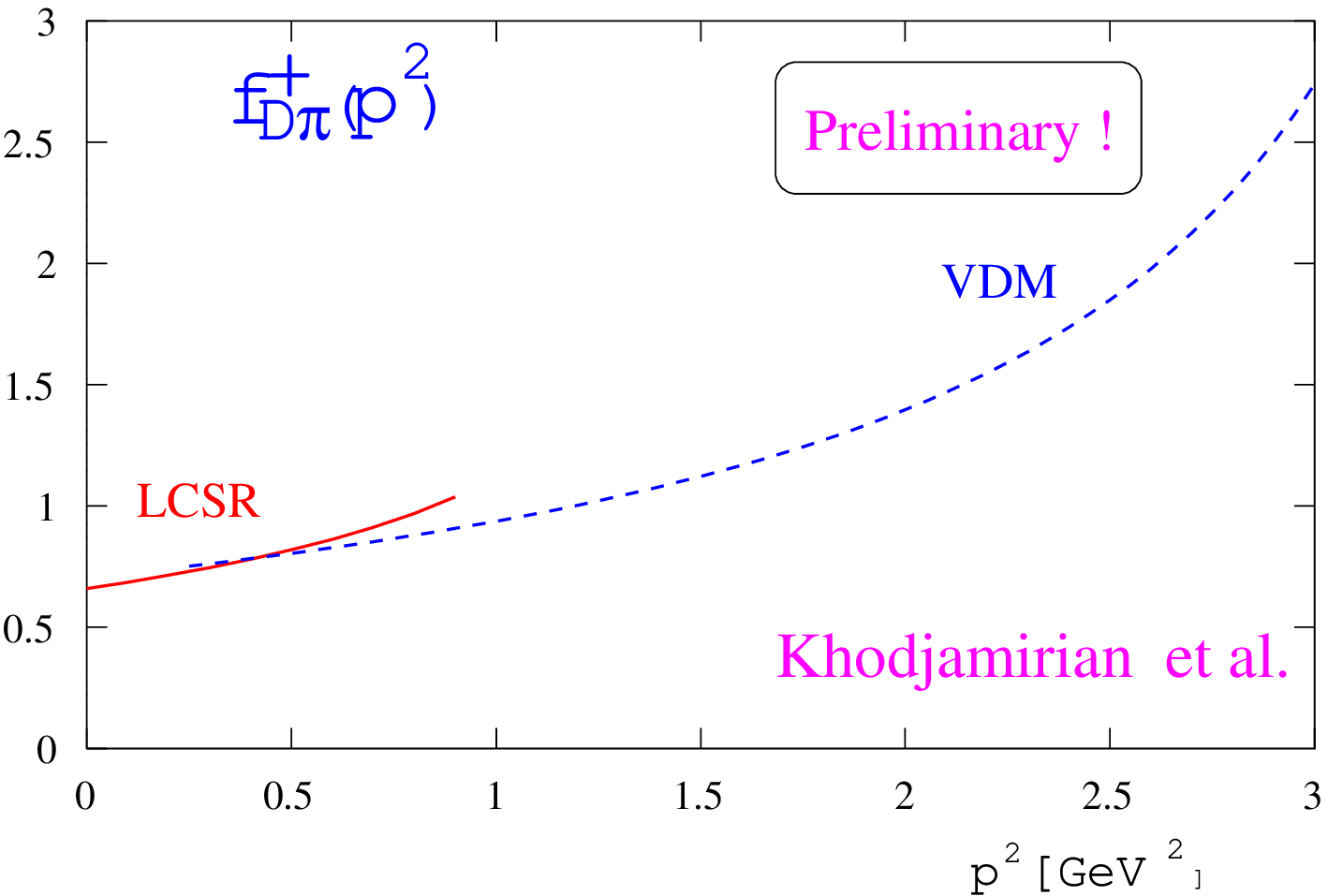,width=7cm}%
        \caption{Form factor $f_+^{D\to\pi}$ from LCSR \cite{K}.} 
	\label{fplusDpi}}

The LCSR result is shown by the solid curve. It is valid
at small $q^2<1$~GeV$^2$  and is matched at higher $q^2$ to the vector
dominance prediction using a yet another new LCSR estimate \cite{KRWY99}
\begin{equation}
  f_{D^*} g_{DD^*\pi} = 2.84 \pm 0.6~\mbox{GeV}^2\,. 
\end{equation}
In the same work, the ratio of decay widths $D\to K$ and $D\to \pi$ is 
calculated and claimed to be strongly sensitive to the value of the strange
quark mass:
\begin{eqnarray}
f_+^{D\to\pi}(0) &=& 0.66 \pm 0.09\,,
\\
f_+^{D\to K}(0) &=&
\left\{
\begin{array}{l}
 0.98\pm 0.11,~~m_s= 100~{\rm MeV}\\
 0.77\pm 0.11,~~m_s= 150~{\rm MeV} \\
 0.66\pm 0.09,~~m_s= 200~{\rm MeV}\nonumber
\end{array}
 \right.
\end{eqnarray}
The scalar form factor $f_0^{D\to\pi}$ was considered in \cite{KRW98}
using LCSR. At $q^2=0$ the result is 
\begin{equation}
  f_0^{D\to\pi}(0) = f_+^{D\to\pi}(0) = 0.68~\mbox{\protect{\cite{KRW98}}}\,
\end{equation}
with, probably, 20\% error.

\section{Semileptonic and rare B decays}

These decays typically involve  a large momentum transfer to
the light hadron in the final state and have to be treated with care.
The corresponding technique, known as light-cone sum rules, was
first suggested in \cite{BBK89,BF89,CZ90} and was developing 
rapidly during past few years. In essence, it presents
a generalization of the classical mean-field type SVZ approach to 
expansion of correlation functions in rapidly varying 
(large momentum) background meson fields. Technically, the difference is
that the short-distance expansion in operators of increasing dimension
is replaced by the light-cone expansion in  increasing twist. 
On this way, vacuum averages of quark and gluon fields (condensates) 
do not appear and are replaced by light-cone meson distribution 
amplitudes, see \cite{Braun98,KR98} for the two recent expositions.   
Premium for this rearrangement is that the approach becomes explicitly 
consistent with the heavy quark expansion.
A detailed comparison of light-cone and conventional QCD sum rules 
can be found in \cite{BB97}.

The decay $B\to\pi e \nu$ was considered many times by different authors
using both conventional \cite{BBD91a,Narison92,CS94} and LCSRs  
\cite{BKR93,BBKR95,KRW98}. Apart from phenomenological importance, this 
decay serves as a convenient test ground for various versions of sum rules
and other theoretical ideas. The new results which I want to report
on this meeting, are obtained using LCSR calculations with NLO accuracy, 
including radiative corrections \cite{KRWY97,BBB97,Ball98}, see 
Fig.~\ref{Bpi1}.  
\FIGURE{\epsfig{file=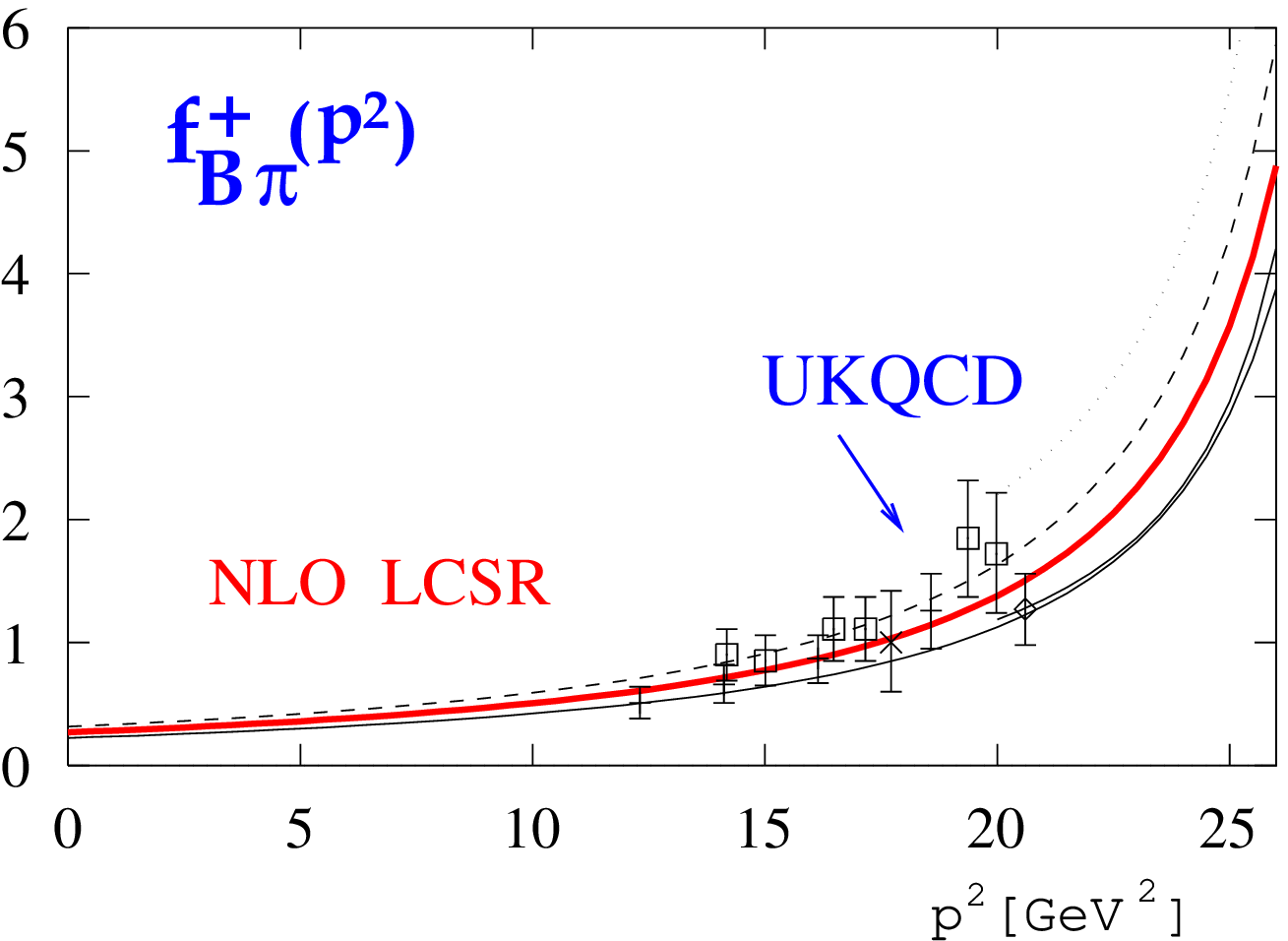,width=7cm}%
        \caption{Form factor $f_+^{B\to\pi}$ from LCSR \cite{K}.} 
	\label{Bpi1}}
At $q^2=0$ one obtains \cite{KRWY97,BBB97,Ball98}
\begin{eqnarray}
   f_+^{B\to\pi} &=& 0.27 \pm 0.05\,,
\end{eqnarray}
where the largest uncertainty comes from pion distribution amplitude.
The calculation automatically satisfies  the unitarity constraints 
\cite{BG97}, as illustrated  in Fig.~\ref{Bpi2}. It should be mentioned
that the LCSR approach is only justified sufficiently far 
from zero recoil, as indicated by arrows in in Fig.~\ref{Bpi2},
and in this  region  cannot violate any  unitarity bounds.    
\FIGURE{\epsfig{file=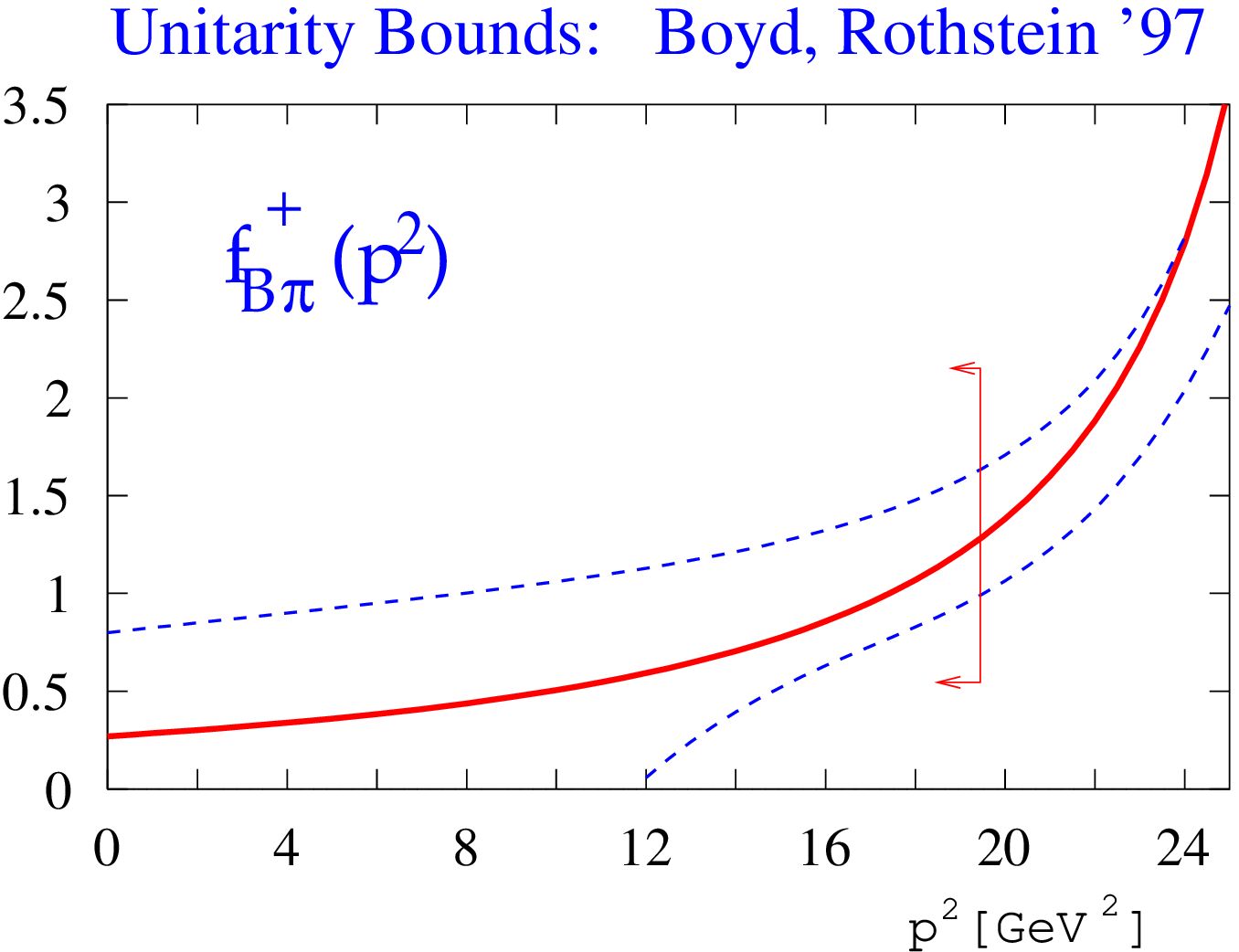,width=7cm}%
        \caption{Form factor $f_+^{B\to\pi}$ from LCSR (solid curve)
         compared with the unitarity bounds (dashes).
         Figure taken from \cite{K}.} 
	\label{Bpi2}}
My personal opinion is that unitarity constraints are not sufficiently 
restrictive and for small $q^2$ do not provide any additional knowledge 
(for heavy-to-light decays) compared  to ``reasonable'' model calculations.  

Many new results were obtained recently for B-decays into light
vector mesons. Early LCSR calculations \cite{ABS94,AOS97,BB97} have been 
updated in \cite{BB98}  by taking into account radiative corrections, 
meson mass corrections  and contributions of meson wave functions of 
higher twist \cite{TW4}.

The results  are presented in 
Fig.~\ref{Kstar} for the particular case $B\to K^*$. Note the error bands
and good agreement with lattice calculations by the UKQCD collaboration
\cite{UKQCD}. Several simple parametrizations of the $q^2$-dependence 
are available, see e.g. \cite{BB98,BK99}.
The values of form factors at $q^2=0$ are compiled in 
Table~2. Note general agreement between different LCSR calculations
and also with lattice results. 
For a detailed analysis of the phenomenological implications of these
results see \cite{ABHH99}. 

\FIGURE{\epsfig{file=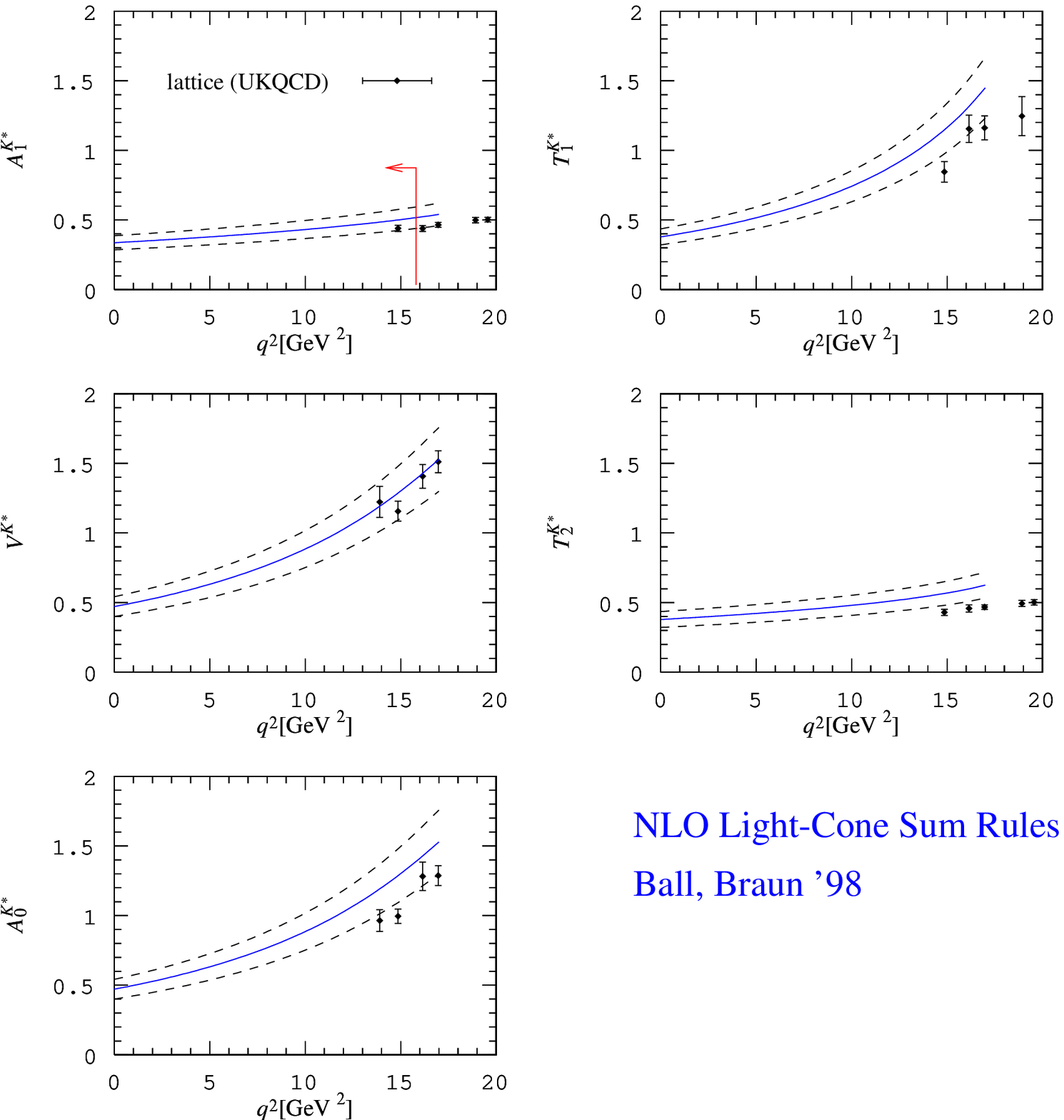,width=13cm}%
        \caption{Rare decay  $B\to K^*$ form factors from LCSR \cite{BB98}.} 
	\label{Kstar}}
\TABLE[ht]{
\renewcommand{\arraystretch}{1.3}
\addtolength{\arraycolsep}{1pt}
$
\begin{array}{|l|cccc|}\hline
& {\rm{ BB98} \cite{BB98}} & {\rm ABS \cite{ABS94}, BB97 \cite{BB97}} &
{\rm AOS \cite{AOS97}}& {\rm UKQCD \cite{UKQCD}}
    \\[-7pt] 
 & {\rm { (NLO~LCSR)}} & {\rm(LCSR)} & {\rm(LCSR)} & {\rm (Lattice+LCSR)} \\ \hline
A_1^\rho(0)  & 0.26\pm0.04 & 0.27\pm0.05 & 0.30\pm0.05 &
0.27^{+0.05}_{-0.04}\\
A_2^\rho(0)  & 0.22\pm0.03 & 0.28\pm0.05 & 0.33\pm0.05 &
0.26^{+0.05}_{-0.03}\\
V^\rho(0)    & 0.34\pm0.05 & 0.35\pm0.07 & 0.37\pm0.07 &
0.35^{+0.06}_{-0.05} \\
T_1^\rho(0)  & 0.29\pm0.04 & 0.24\pm0.07 & 0.30\pm0.10 & - \\
T_3^\rho(0)  & 0.20\pm0.03 & -          & 0.20\pm0.10 & - \\
A_1^{K^*}(0) & 0.34\pm0.05 & 0.32\pm0.06 & 0.36\pm0.05 &
0.29^{+0.04}_{-0.03} \\
A_2^{K^*}(0) & 0.28\pm0.04 & -          & 0.40\pm0.05 & - \\
V^{K^*}(0)   & 0.46\pm0.07 & 0.38\pm0.08 & 0.45\pm0.08 & - \\
T_1^{K^*}(0) & 0.38\pm0.06 & 0.32\pm0.05 & 0.34\pm0.10 &
0.32^{+0.04}_{-0.02} \\
T_3^{K^*}(0) & 0.26\pm0.04 & -          & 0.26\pm0.10 & - \\
\hline
\end{array}
$
\renewcommand{\arraystretch}{1}
\addtolength{\arraycolsep}{-1pt}
\caption{
Comparison of results from different works on $B\to\rho,K^*$ form factors
at $q^2=0$ \cite{BB98}.}
}

The b-quark mass dependence of the form factors deserves a special
discussion. As observed in \cite{CZ90,ABS94} all form factors in question
have a universal behaviour  $\sim 1/m_b^{3/2}$ in the heavy quark limit.
This scaling law is suggested by the analysis of leading integration
regions in Feynman diagrams (see discussion in \cite{BB97,Braun98}) and is 
valid for both ``soft'' and ``hard'' contributions separately.
To be precise, the scaling is $\sim \sqrt{m_b}/E^2$ 
\cite{Braun98,BB97,CYOPR98}
where $E$ is the light-meson energy in the final state 
(in the B-meson rest frame). The factor $1/E^2$ arises invariantly either as
gluon virtuality in hard rescattering, or from the overlap integral 
between soft wave functions, while $\sqrt{m_b}$ comes from the B-meson
wave function normalization. This scaling law is supported  by LC sum rules
which also suggest that the soft contribution is dominant for realistic 
b-quark mass values. If the hard rescattering correction is neglected,
several symmetry relations can be derived between semileptonic and rare
decay form factors in the heavy quark limit, see \cite{CYOPR98}.
The LCSR calculations are explicitly consistent with these relations 
and in fact suggest that the corrections are numerically rather small 
\cite{ABS94,BB98,CYOPR98}.
On the other hand, the b-quark mass dependence comes out to be rather 
far from the asymptotic scaling limit. Rough estimates \cite{ABS94,KRW98}
suggest large preasymptotic corrections of order
\begin{equation}
  F(q^2=0) \sim {m_b^{-3/2}} F_{\rm stat}
\left[1-\frac{C}{m_b}+\ldots\right]
\end{equation} 
with $C\sim 1 - 1.5$~GeV. 

As far as technical implementation is concerned, the existing LCSR 
calculations present the state of the art of this approach and 
it will be difficult to improve them in near future. They can be updated
if new information about light-cone meson distribution functions 
becomes available. There exist, however, a few issues which still have
to be clarified. First, LC sum rules systematically indicate larger SU(3)
breaking corrections compared to lattice calculations, and the origin
of this is not understood. Second, in the treatment of sum rules 
themselves one probably can improve on the treatment of kinematical 
factors (separation of Lorentz structures and invariant amplitudes),
 in particular, going  over to calculations of helicity amplitudes.
 
\section{$g_{B^*B\pi}$, $g_{D^*D\pi}$ etc.}      

A detailed study of the pion coupling to heavy mesons was carried out 
in \cite{BBKR95} using both light-cone and conventional 
 sum rules with the result
\begin{eqnarray}
     g_{B^*B\pi} &=& 29\pm 3\,,
\nonumber\\
      g_{B^*B\pi} &=& 12.5\pm 1\,,
\label{coupl}
\end{eqnarray}
where the errors are probably somewhat underestimated.  
In the heavy quark limit one obtains \cite{BBKR95}
\begin{eqnarray}
 g_{B^*B\pi} &=& 
    \frac{2m_B}{f_\pi}\,{\hat g}\left(1+\frac{{\Delta}}{m_B}\right),
\nonumber\\
 \hat g &=& 0.32\pm 0.02\,,
 \nonumber\\
  \Delta &=& (0.7 \pm 0.1)\,{\rm GeV}\,. 
\end{eqnarray}
Very recently, radiative corrections to these sum rules 
have been calculated \cite{KRWY99}. They turn out to be 
negative and decrease the values of the couplings by approximately 20\%.
To my opinion this reduction has to be taken with some caution since 
the main input parameter in this sum rule -- the pion distribution amplitude 
in the middle point -- was determined in \cite{BF89} without taking 
radiative corrections into account. The analysis of \cite{BF89} is in fact 
quite old and its update is long overdue.

Similar sum rules (to leading order only) have been derived for 
decay constants $g_{B^*B\rho}$ and $g_{D^*D\rho}$ \cite{ADIP96}, 
for radiative $B^*\to B\gamma $ decays \cite{AIP97} and for couplings to
positive parity heavy mesons \cite{APS96,CdF97}.

\section{Other topics in B-decays}

Concluding this part, I want to mention two other applications of sum rules
which I find interesting.

First,  an exploratory study of 
nonfactorizable contributions to the $B\to J/\Psi K$ decay \cite{KR98,KR98a}. 
In the usual factorisation approximation the corresponding matrix element
is proportional to the sum $a_2 = c_2(m_b)+\frac13 c_1(m_b) \simeq 0.155$,
where $c_1$ and $c_2$ are the coefficients in the effective Lagrangian
in common notation. The experimental data are usually analysed 
introducing an effective coefficient $a_{\rm 2,eff}^{B\psi K}$
to take into account nonfactorizable effects and the current result 
$|a_{\rm 2, exp}^{B \psi K}| = 0.31\pm 0.02$ can be interpreted as 
strong violation of the factorization. The effective coefficient 
$a_2$ was estimated in QCD sum rules \cite{KR98,KR98a} using both 
the standard technique \cite{BS87} developed originally for D-decays,
and the LCSRs. Numerical results are still very unstable and preliminary,
but, interestingly enough, the sum rules seem to indicate a {\em negative}
value for $a_2$. Although calculations are difficult, potentially 
this is a large and interesting field of applications. 

Second, the LCSR approach can be and was already used to estimate 
long-distance contributions
in decays like $B\to K^*\gamma, B\to \rho^*\gamma$, see 
\cite{KSW95,AB95,KRSW97}. Most importantly, such calculations 
predict the relative sign of short- and long-distance amplitudes
and, therefore, their interference. 
The existing results can be improved and 
extended to $B\to K^* \ell \bar \ell$ and similar transitions. 
The sum rules of this type are rather well understood and can be expected 
to give reliable predictions.

\section{Heavy baryons} 

Studies of heavy baryons in QCD sum rules are in general difficult. 
They have been going on for some time, but, to my opinion, the results 
still are rather preliminary. I will discuss two applications 
to problems of direct phenomenological relevance.

First is the calculation of baryon matrix elements of four-fermion 
operators \cite{CdF96} that are responsible for the $1/m_b^2$ corrections
to life-time differences between beautiful baryons and mesons. 
The corresponding sum rules are very unstable because of high dimension
and can only produce an order of magnitude  estimate. The result of the 
calculation is negative: 
Despite all uncertainties one can claim that the relevant matrix elements 
cannot be as large as to  explain the observed ratio 
$\tau(\Lambda_d)/\tau(B_d)$. 

Second, several QCD sum rule calculations exist of the semileptonic decays 
of heavy baryons
$\Lambda_b\to\Lambda_c \ell \bar\nu$, $\Lambda_c\to\Lambda \ell \bar\nu$
 and $\Lambda_b\to p \ell \bar\nu$, 
 \cite{DHHL96,HQY98,Dosch99}, see Fig.~\ref{dosch}, and also of rare decays 
$\Lambda_b\to\Lambda\gamma$, $\Lambda_b\to\Lambda \ell \bar \ell$ 
\cite{HY99}. 
\FIGURE{\epsfig{file=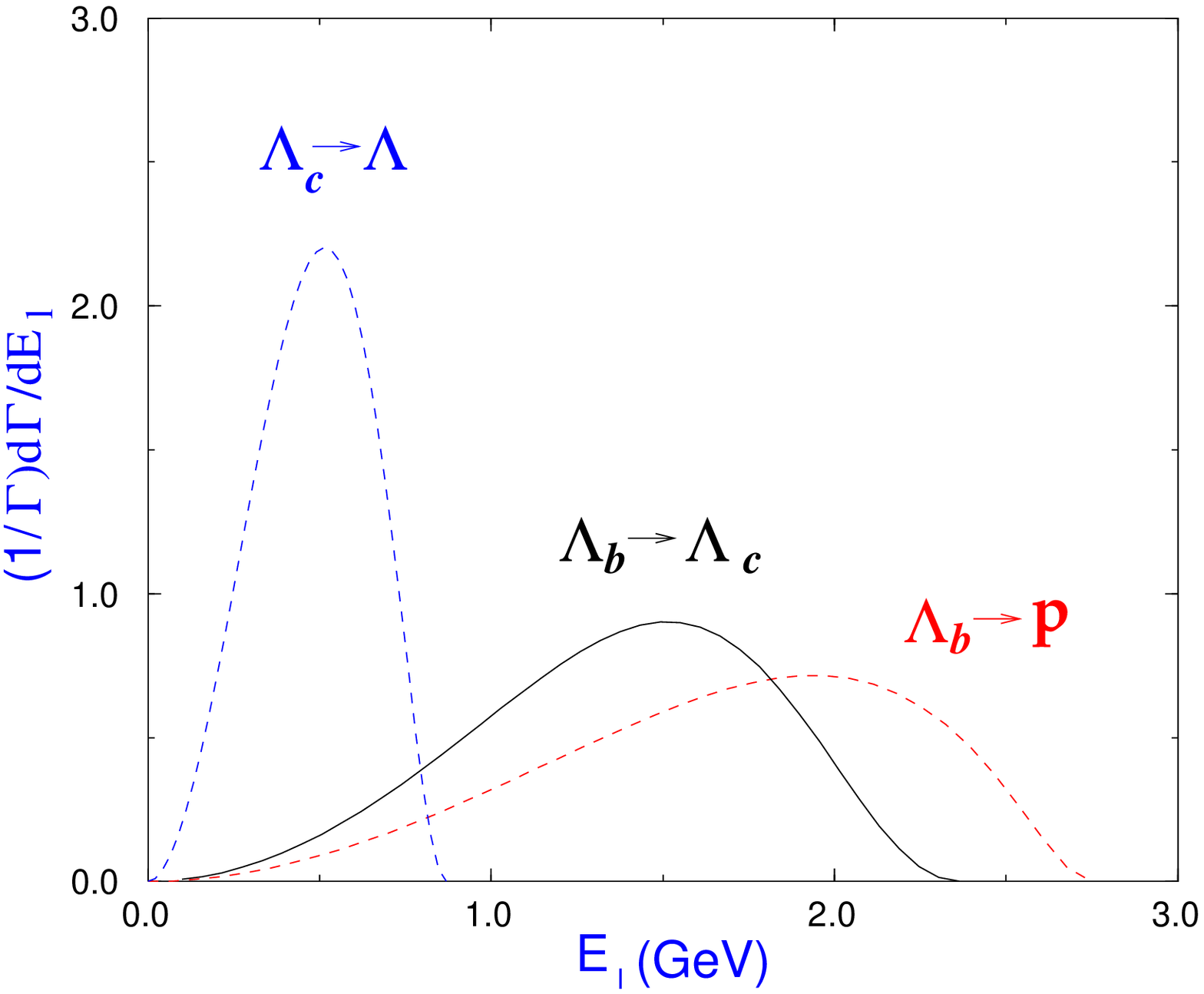,width=7cm}%
        \caption{Lepton energy spectra in decays of heavy baryons 
       \cite{Dosch99}.} 
	\label{dosch}}
For the decay widths one obtains, e.g. \cite{Dosch99}
\begin{equation}
\Gamma(\Lambda_b\to p\,\ell^- \bar \nu_\ell) =
(1.7\pm 0.7)\times 10^{-11}|V_{ub}|^2 \,\mbox{\rm GeV} 
\end{equation}
All these studies are carried out using traditional three-point sum rules. 
The LCSR technology has not been applied yet because of almost complete
absence of information on higher-twist baryon wave functions.
In the case of a light baryon in the final state, the same critisism
applies, therefore,  to this calculations as for heavy-to-light meson 
transitions \cite{BB97}.

\section{Conclusions}

The main new development in the QCD sum rules during the past couple of 
years have been, to my opinion, the arrival of a new generation
of NLO LCSR calculations, incorporating radiative corrections and 
new results on meson wave functions of leading and higher twists. 
Detailed state-of-the-art studies are now available using such sum rules
for the form factors of heavy-to-light transitions at large recoil.

Among the problems where further developments can be expected and the sum
rules are likely to make an impact, I can mention:
\begin{itemize}
\item{} Light-cone sum rules for heavy baryons. Dedicated studies 
        of higher-twist baryon distribution amplitudes beyond leading twist
        are, however, required as a first step.
\item{} Long-distance contributions to various rare decays.
\item{} Nonfactorizable contributions to nonleptonic decays like 
        $B\to\pi\pi$ etc.
\item{} Matrix elements of penguin operators.
\end{itemize}

Interesting problems where exploratory studies are neccessary to 
determine feasibility of the QCD sum rule calculations are, 
among others: Colour-octet matrix elements in charmonium, parton distributions
in B mesons and $B_c$ decays (see \cite{KLO99}). 
Further progress can eventually be achieved
in cooperation of sum rules with lattice calculations (e.g. to 
determine meson distribution amplitudes) and by studying exclusive processes
with light hadrons at comparable momentum transfers.

\acknowledgments
I am grateful to A. Ali, E. Bagan, I. Balitsky, P. Ball, H.G. Dosch, 
I. Halperin, A. Khodjamirian, 
R. R\"uckl and H. Simma for collaboration on the topics covered in 
this report. Special thanks are due to A. Khodjamirian and R. R\"uckl 
for the possibility to use the results of \cite{K} prior publication.



\begin{thebibliography}{999}
\bibitem{SVZ}
M.A. Shifman, A.I. Vainstein and V.I. Zakharov, \npb{147}{1979}{385}.
%
\bibitem{Shuryak82}
E.V.~Shuryak, \npb{198}{1982}{83}; \plb{93}{1980}{134}.
%
\bibitem{AE83}
T.M.~Aliev and V.L.~Eletsky, \sjnp{38}{1983}{936}.
%
\bibitem{scatter}
C.A. Dominguez and N. Paver, \plb{197}{1987}{423};
                             \ibid{199}{1987}{596 (E)};\\
L.J. Reinders, \prd{38}{1988}{423};\\
S. Narison, \plb{198}{1987}{104};\\
P. Colangelo {\em et al.}, \plb{269}{1991}{201}.
%
\bibitem{BG92}
D.J. Broadhurst and A.G. Grozin, \plb{274}{1992}{421}.
%
\bibitem{BBBD92}
E. Bagan {\em et al.}, \plb{278}{1992}{457}.
%
\bibitem{N92}
M. Neubert, \prd{45}{1992}{2451}.
%
\bibitem{BenekeHF8}
M. Beneke, these proceedings.
%
\bibitem{BSS99rep}
M. Beneke, A. Signer and V.A. Smirnov, 
{\em A Two loop application of the threshold expansion: 
The Bottom quark mass from $b \bar b$ production},
\hepph{9906476}.
%
\bibitem{BS99}
M. Beneke and A. Signer, {\em The Bottom MS-bar quark mass from sum rules at 
next-to-next-to-leading order}, \hepph{9906475}.
%
\bibitem{PP98}
A.A. Penin and A.A. Pivovarov, 
\npb{549}{1999}{217}; \plb{435}{1998}{413}.
%
\bibitem{H98}
A. Hoang, \prd{59}{1999}{014039}.
%
\bibitem{MY98}
K. Melnikov and A. Yelkhovsky, \prd{59}{1999}{114009}.
%
\bibitem{JP97/98}
M. Jamin and A. Pich, \npb{507}{1997}{334}; \hepph{9810259}.
%
\bibitem{PY98}
A. Pineda and F.J. Yndurain, 
{\em Comment on 'Calculation of quarkonium spectrum and 
$m_b$, $m_c$ to order $\alpha_s^4$'},
\hepph{9812371}.
%
%
\bibitem{GGRM99}
V. Gimenez {\em et. al}, {\em  NNLO unquenched calculation 
of the b quark mass}, \heplat{9909138}.
%
\bibitem{BBB95}
P. Ball, M. Beneke and V. Braun, \prd{52}{1995}{3929}.
%
\bibitem{Dominguez93}
C.A. Dominguez, 
in {\em Proc. of the Third Workshop on the Tau-Charm Factory},
Marbella, Spain (1993), Ed. J. Kirkby and R Kirkby,
Editions Fronti\'eres, p.357.
%
\bibitem{Babar}
{\em The BaBar Physics Book, Appendix D}, 
Ed.: P.F. Harrison and H.R. Quinn, 
SLAC-R-504 (1998). 
%
\bibitem{BBKR95}
V.M. Belyaev {\em et al.}, \prd{51}{1995}{6177}.
%
\bibitem{KRWY99}
A. Khodjamirian {\em et al.}, \plb{457}{1999}{245}.
%
\bibitem{ES92}
V. Eletsky and E. Shuryak, \plb{276}{1992}{191}.
%
\bibitem{Ball94}
P. Ball, \npb{421}{1994}{593}.
%
\bibitem{Nreview}
M. Neubert, \prep{245}{1994}{259}.
%
\bibitem{BB94}
P. Ball and V.M. Braun \prd{49}{1994}{2472}.
%
\bibitem{N96}
M. Neubert, \plb{389}{1996}{727}.
%
\bibitem{BSU97}
I. Bigi, M. Shifman and  N. Uraltsev, \arnps{47}{1997}{591}. 
%
\bibitem{BL98}
B. Blok and M. Lublinsky, \prd{57}{1998}{2676}; \ibid{58}{1998}{019903(E)}.
%
\bibitem{OS89}
 A.A. Ovchinnikov and  V.A. Slobodenyuk, \sjnp{50}{1989}{891}.
%
\bibitem{BG90}
V.N. Baier and A.G. Grozin, \zpc{47}{1990}{669}.
%
\bibitem{CNOP91}
P. Colangelo {\em et al.}, \plb{269}{1991}{201}.
%
\bibitem{Rad91}
A.V. Radyushkin, \plb{271}{1991}{218}.
%
\bibitem{Ball92}
P. Ball, \plb{281}{1992}{133}.
%
\bibitem{BS93}
B. Blok and M. Shifman, \prd{47}{1993}{2949}.
%
\bibitem{IWradcor}
E. Bagan, P. Ball and P. Gosdzinsky, \plb{301}{1993}{249};\\
M. Neubert, \prd{47}{1993}{4063}.
%
\bibitem{CNP92}
P. Colangelo, G. Nardulli and N. Paver, \plb{293}{1992}{207}.
%
\bibitem{CFP98}
P. Colangelo, F. De Fazio and N. Paver, \prd{58}{1998}{116005}; 
\hepph{9810478}.
%
\bibitem{sub}
M. Neubert, \prd{46}{1992}{3914};\\
M. Neubert, Z. Ligeti and Y. Nir, \plb{301}{1993}{101};
\prd{47}{1993}{5060}.
%
\bibitem{HLD99}
M. Huang, C. Li and Y. Dai, {\em
QCD sum rule analysis of the subleading Isgur-Wise form-factor 
$\tau_1(vv')$ and $\tau_2(vv')$  for $B\to D_1 \ell \bar \nu$ and
$B\to D_2^\ast \ell \bar \nu$}, \hepph{9909307}. 
%
\bibitem{AEK84}
T.M. Aliev, V.L. Eletsky and Ya.I. Kogan, \sjnp{40}{1984}{527}.
%
\bibitem{AOS89}
T.M. Aliev, A.A. Ovchinnikov and V.A. Slobodenyuk,
Preprint IC/89/382 (unpublished).
%
\bibitem{BBDN91}
P. Ball {\em et al.}, \plb{259}{1991}{481}.

\bibitem{BBD91}
P. Ball, V.M. Braun and  H.G. Dosch, \prd{44}{1991}{3567}.
%
\bibitem{Ball93}
P. Ball, \prd{48}{1993}{3190}.
%
\bibitem{K}
A.~Khodjamirian, R.~R\"uckl, S. Weinzierl, 
C.W.~Winhart and O.~Yakovlev, {\em 
Predictions on $B \to \pi \bar{l} \nu $, $D \to \pi \bar{l} \nu $ 
and $D\to K \bar{l} \nu $ from QCD Light-Cone 
Sum Rules}, Preprint WUE-ITP-99-017, paper in preparation.
%
\bibitem{KRW98}
A. Khodjamirian, R. Ruckl and C.W. Winhart, \prd{58}{1998}{054013}.
%
\bibitem{BBK89}
I.I. Balitsky, V.M. Braun and A.V. Ko\-les\-ni\-chen\-ko, 
\sjnp{44}{1986}{1028}; \yf{48}{1988}{855}; \sjnp{48}{1988}{348};
\npb{312}{1989}{509}.
%
\bibitem{BF89}
V.M. Braun and I.E. Filyanov, \zpc{44}{1989}{157}.
%
\bibitem{CZ90}
V.L. Chernyak and I.R. Zhitnitsky, \npb{345}{1990}{137}.
%
\bibitem{Braun98}
V.M. Braun, {\em  Light cone sum rules}, 
\hepph{9801222}.
%
\bibitem{KR98}
A. Khodjamirian and R. R\"uckl, 
{\em QCD sum rules for exclusive decays of heavy mesons},
\hepph{9801443}.
%
\bibitem{BB97}
P. Ball and V.M. Braun, \prd{55}{1997}{5561}. 
%
\bibitem{BBD91a}
P. Ball, V.M. Braun and  H.G. Dosch, \plb{273}{1991}{316}.
%
\bibitem{Narison92}
S. Narison, \plb{283}{1992}{384}
%
\bibitem{CS94}
P. Colangelo and P. Santorelli, \plb{327}{1994}{123}.
%
\bibitem{BKR93}
V.M. Belyaev, A. Khodjamirian and R. R\"uckl, \zpc{60}{1993}{349}.
%
\bibitem{KRWY97}
A. Khodjamirian {\em et al.}, \plb{410}{1997}{275}.
%
\bibitem{BBB97}
E. Bagan, P. Ball and V.M. Braun, \plb{417}{1998}{154}.
%
\bibitem{Ball98}
P. Ball, \jhep{09}{1998}{005}.
%
\bibitem{BG97}
C.G. Boyd and I.Z. Rothstein, \plb{420}{1998}{350}.
%
\bibitem{ABS94}
A. Ali, V.M. Braun and H. Simma, \zpc{63}{1994}{437}.
%
\bibitem{AOS97}
T.M. Aliev, A. Ozpineci and M. Savci, \prd{56}{1997}{4260}.
%
\bibitem{BB98}
P. Ball and V.M. Braun, \prd{58}{1998}{094016}.
%
\bibitem{TW4}
P. Ball {\em et al.} \npb{529}{1998}{323};\\
P. Ball and V.M Braun, \npb{543}{1999}{201}.
%
\bibitem{BK99}
D. Becirevic and A.B. Kaidalov, {\em
Comment on the heavy $\to$ light form-factors}, \hepph{9904490}. 
%
\bibitem{UKQCD}
UKQCD Collaboration, L. Del Debbio {\em et al.}, \plb{416}{1998}{392}.
%
\bibitem{ABHH99}
A. Ali {\em et al.}, {\em
A Comparative study of the decays $B\to (K, K*) \ell^+ \ell^-$ in 
standard model and supersymmetric theories}, \hepph{9910221}. 
%
\bibitem{CYOPR98}
J. Charles {\em et al.}, \prd{60}{1999}{014001}.
%
\bibitem{ADIP96}
T.M. Aliev {\em et al.}, \prd{53}{1996}{355}.
%
\bibitem{AIP97}
T.M. Aliev, E. Iltan and  N.K. Pak, \zpc{73}{1997}{293}.
%
\bibitem{APS96}
T.M. Aliev, N.K. Pak and M. Savci, \plb{390}{1997}{335}.
%
\bibitem{CdF97}
P. Colangelo and F. De Fazio, {\em Eur.Phys.J.} {\bf C4} (1998) 503.
%
\bibitem{KR98a}
 A. Khodjamirian and R. R\"uckl, {\em 
Exclusive Nonleptonic Decays of Heavy Mesons in QCD}, \hepph{9807495}.
%
\bibitem{BS87}
B.Yu. Blok and M.A. Shifman, \sjnp{45}{1987}{135,301,522}.
%
\bibitem{KSW95}
A. Khodjamirian, G. Stoll and D. Wyler, \plb{358}{1995}{129}.
%
\bibitem{AB95}
A. Ali and V.M. Braun, \plb{359}{1995}{129}. 
%
\bibitem{KRSW97}
A. Khodjamirian {\em et al.}, \plb{402}{1997}{167}.
%
\bibitem{CdF96}
P. Colangelo and F. De Fazio, \plb{387}{1996}{371}.
%
\bibitem{DHHL96}
Y.-B. Dai {\em et al.}, \plb{387}{1996}{379}.
%
\bibitem{HQY98}
C.-S. Huang, C.-F. Qiao and  H.-G. Yan, \plb{437}{1998}{403}.
%
\bibitem{Dosch99}
R.S.M. de Carvalho {\em et al.}, \prd{60}{1999}{034009}.
%
\bibitem{HY99}
C.-S. Huang and H.-G. Yan, \prd{59}{1999}{114022}.
%
\bibitem{KLO99}
V.V. Kiselev, A.A. Likhoded and A.I. Onishchenko, {\em 
Semileptonic $B_c$ meson decays in sum rules of QCD and NRQCD}, 
\hepph{9905359}. 

\end{thebibliography}
\end{document}